\documentclass[preprint,12pt]{elsarticle}

\usepackage{amssymb}
\usepackage{subfigure}
\usepackage{bm,amsmath}
\usepackage{graphicx}

\begin{document}

\begin{frontmatter}



\title{\textbf{A semi-Lagrangian gas-kinetic scheme for smooth flows}}


\author[hua]{Peng Wang}
\ead{sklccwangpeng@hust.edu.cn}
\author[hua,hub]{Zhaoli Guo \corref{cor}}
 \cortext[cor]{Corresponding author}
 \ead{zlguo@hust.edu.cn}
\address[hua]{State Key Laboratory of Coal Combustion, Huazhong
University of Science and Technology\\
\small Wuhan  430074  \ P.R.China}
\address[hub]{Beijing Computational Science Research Center\\
 \small Beijing, 100084 \ P.R.China }

\begin{abstract}

In this paper, a semi-Lagrangian gas-kinetic scheme is developed for smooth flows based on the Bhatnagar-Gross-Krook (BGK) equation. As a finite-volume scheme, the evolution of the average flow variables in a control volume is under the Eulerian framework, whereas the construction of the numerical flux across the cell interface comes from the Lagrangian perspective. The adoption of the Lagrangian aspect makes the collision and the transport mechanisms intrinsically coupled together in the flux evaluation. As a result,  the time step is independent of the particle collision time and solely determined by the Courant-Friedrichs-Lewy (CFL) conditions. A set of simulations are carried out to validate the performance of the new scheme. The results show that with second-order spatial accuracy, the scheme exhibits low numerical dissipation, and can accurately capture the Navier-Stokers solutions for the smooth flows with viscous heat dissipation from the low-speed incompressible to hypersonic compressible regimes.

\end{abstract}


\end{frontmatter}


\section{Introduction}
\label{intro}

In recent years,  kinetic methods have drawn particular attention as newly-developing computational fluid dynamics(CFD) technology. Unlike the conventional CFD methods based on direct discretizations of the Navier-Stokes (NS) equations, kinetic methods are based on kinetic theory or  micropartical dynamics, which provides theoretical connection between hydrodynamics and the underlying microscopic physics, and thus yields efficient tools for multiscale flows. Up to date, a variety of mesoscopic methods have been proposed, such as the lattice gas cellular automata (LGCA) \cite{LGA}, the lattice Boltzmann equation (LBE) \cite{LBM,Guobook}, the gas-kinetic scheme (GKS) \cite{Xu_Prandtl,ugks,Mandal1994,FVS,KFVS,DGKS}, and the smoothed particle hydrodynamics(SPH)\cite{SSH}, among which the LBE and GKS are specifically designed for CFD. The kinetic nature of the LBE and GKS has led to many distinctive advantages that distinguish them from the classical CFD methods. Particularly, the GKS for the Navier-Stokes solutions has been well developed \cite{Xu_Prandtl,FVS,KFVS,Xu_multidimetion,tao,guo2008,CHEN,TANG} , and successfully applied to a variety of flow problems \cite{hypersonic,ugks2,rarefied,micro,multimaterial,reactive}.

As a kind of finite-volume scheme, the key ingredient in GKS for NS solutions is to construct the flux at the cell interface. With different approaches, several kinetic schemes have been developed based on the kinetic theory, such as the  Kinetic Flux Vector Splitting (KFVS) \cite{FVS,KFVS,Warming1981} scheme based on the collisionless Boltzmann equation and the GKS based on the Bhatnagar-Gross-Krook (BGK) equation where the particle collisions are considered in the construction of the numerical flux. It is shown that the GKS methods avoids the ambiguity of adding \emph{ad hoc} ``collisions" for the KFVS to reduce the numerical dissipations \cite{FVS,KFVS}. Among the BGK-type schemes, the gas-kinetic BGK-NS scheme for the NS solutions has been well developed \cite{Xu_Prandtl}, and has been successfully applied for the continuum flow simulation from low-speed incompressible to hypersonic compressible flows \cite{Xu_multidimetion,hypersonic}.

In this paper, we present a semi-Lagrangian gas-kinetic scheme (SLGKS), as an alternative BGK-type scheme, for smooth flows . The most distinguished feature of the proposed scheme is that the construction of the flux at the cell interface is based on the discrete characteristic solution of the BGK equation, which comes from the Lagrangian aspect. This approach results in the particle collision and transport mechanisms coupled together within a time step, which makes the new scheme exhibited very low numerical dissipation and the time step decoupled from the particle collision time. In order to validate present scheme to be a feasible NS solver in the under-resolved region, a set of simulations are carried out, including the thermal Couette flow, thermal Poiseuille flow, the shock structure problem and the laminar flow over a flat plate.

The rest of the article is organized as follows. In Sec.~\ref{SLGKS}, the semi-Lagrangian gas-kinetic scheme is developed. Numerical tests are made in Sec.~\ref{numerical} to validate the performance of the new scheme, and finally some conclusions are drawn in Sec.~\ref{con}.

 \section{Semi-Lagrangian gas-kinetic scheme}
 \label{SLGKS}
 The Boltzmann equation expresses the behavior of a many-particle kinetic system in terms of the evolution equation of the singlet gas distribution function. One of its simplified version is the BGK model\cite{BGK},
\begin{equation}\label{BGK}
\frac{\partial f}{\partial t}+{\bm \xi}\cdot\nabla f=\Omega\equiv\frac{g-f}{\tau},
\end{equation}
where $f$ is the gas distribution function and $g$ is the equilibrium state approached by $f$. Both $f$ and $g$ are functions of space $\bm x$, time $t$, particle velocities ${\bm \xi}$, and internal variable $\bm \eta$. The particle collision time $\tau$ is related to the viscosity and
the heat conduction coefficients. The equilibrium state is a Maxwellian distribution,
\begin{equation}\label{MAX}
g=\frac{\rho}{{(2\pi R T)}^{(D+K)/2}}\exp\left(-\frac{({\bm \xi}-{\bm u})^{2}+{\bm \eta}^2}{2RT}\right)
\end{equation}
where $D$ is the spatial dimension, $K$ is the internal degree of freedom, $\rho$ is the density, ${\bm u}$ is the macroscopic velocities,
$R$ is the gas constant, and $T$ is the gas temperature. The connection  between the distribution function $f$ and conservative variable $\bm{W}$ is
\begin{equation}\label{MAX}
\bm{W}=\begin{pmatrix}
\begin{array}{ll}
\rho \\
\rho {\bm u}\\
\rho\epsilon
\end{array}
\end{pmatrix}
=\int{\bm \psi}f d\Xi,
\end{equation}
and the fluxes are computed as,
\begin{equation}\label{flux}
\bm{F}=\begin{pmatrix}
\begin{array}{ll}
F_{\rho} \\
F_{\rho {\bm u}}\\
F_{\rho\epsilon}
\end{array}
\end{pmatrix}
=\int{{\bm\xi}\bm \psi}f d\Xi,
\end{equation}
where $d\Xi=d{\bm \xi}d{\bm\eta}$ is the volume element in phase space with $d\bm \eta=d{\eta}_{1}d{\eta}_{2} \dots d{\eta}_{K}$, and ${\bm\psi}$ is given by
\begin{equation*}
{\bm\psi}=[{\psi}_{1},{\psi}_{2},{\psi}_{3}]^{T}=\left[1,{\bm \xi},\frac{1}{2}\left({\bm \xi}^{2}+{\bm \eta}^{2}\right)\right]^{T},
\end{equation*}
Since mass, momentum and energy are conserved during particle collisions, $f$ and $g$ satisfy the conservation constraint
\begin{equation}
\int(g-f)\bm\psi d\Xi=0,
\end{equation}
at any point in space and time.

In order to develop a finite volume scheme, the computational domain is first divided into a set of control volumes.  Then we multiply $\bm{\psi}$ on both sides of Eq.~\eqref{BGK}, and integrate it in phase space and physical space over a control volume $V_i$ from $t_n$  to $t_{n+1}$, due to the conservation of conservative variables during particle collision process,  the update of the conservative variables at the center of the $V_i$ becomes

\begin{equation}\label{update1}
{{\bm W}_i}^{n+1}={{\bm W}_i}^{n}-\frac{\Delta t}{|{V}_i|}\ {\bm F}^{n+1/2},
\end{equation}
where
\begin{equation}\label{fluxh}
{\bm F}^{n+1/2}=\int {\int}_{\partial V_i}\left({\bm\xi}\cdot{\bm n} \right)\bm \psi f\left(\bm x,t_n+h\right) d{\bm S}d\Xi
 \end{equation}
 is the macroscopic flux across the cell interface and $h=\Delta t/2$. The mid-point rule is employed in the time domain integration of the convection term.

The key ingredient in updating the averaged conserved variables according to Eq.~\eqref{update1} is to evaluate the flux ${\bm F}^{n+1/2}$ , which can be solely determined by the gas distribution function $f\left(\bm x, t_n+h\right)$. Here the Lagrangian perspective is applied in the construction of
$f\left(\bm x,t_{n}+h\right)$: the Eq.~\eqref{BGK} is integrated within a half time step along the characteristic line with the end
point (${\bm x}_b$) located at the cell interface, and the trapezoidal rule is used to evaluate the collision term,
\begin{equation}\label{DB}
f\left({\bm x}_b,{\bm \xi},t_n+h\right)-f\left({\bm x}_b-{\bm \xi}h,{\bm \xi},t_n\right)=\frac{h}{2}\left[\Omega({\bm x}_b,{\bm \xi},t_n+h)+\Omega({\bm x}_b-{\bm \xi}h,{\bm \xi},t_n)\right].
\end{equation}
In order to remove the implicity of  Eq.~\eqref{DB}, we introduce two auxiliary distribution functions
 \begin{subequations}\label{trans1}
 \begin{equation}
 \bar{f}=f-\frac{h}{2}\Omega=\frac{2\tau+h}{2\tau}f-\frac{h}{2\tau}f^{eq},
  \end{equation}
 \begin{equation}
 {\bar{f}}^+=f+\frac{h}{2}\Omega=\frac{2\tau-h}{2\tau}f+\frac{h}{2\tau}f^{eq}.
 \end{equation}
 \end{subequations}
Note that the particle collision effect is included in the above evolution of the interface gas distribution function, this is the key for the success of this kinetic method. Then Eq.~\eqref{DB} can be rewritten as
\begin{equation}\label{DB1}
\bar{f}\left({\bm x}_b,{\bm \xi},t_n+h\right)={\bar{f}}^+\left({\bm x}_b-{\bm \xi}h,{\bm \xi},t_n\right)
\end{equation}
\begin{figure}
\centering
\includegraphics[width=0.6\textwidth]{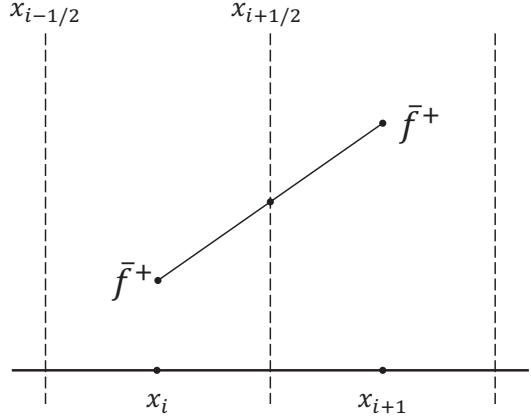}
\caption{Schematic of one-dimensional cell geometry}\label{fig}
\end{figure}
For smooth flows,  ${\bar{f}}^+({\bm x}_b-{\bm \xi}h,{\bm \xi},t_n)$ can be reconstructed by making Taylor expansion around the cell interface ${\bm x}_b$,
\begin{equation}\label{DB2}
{\bar{f}}^+({\bm x}_b-{\bm \xi}h,{\bm \xi},t_n)={\bar{f}}^+({\bm x}_b,{\bm \xi},t_n)-h{\bm \xi}\cdot{\bm \sigma}_b,
\end{equation}
where ${\bm \sigma}_b=\nabla {\bar{f}}^+({\bm x}_b,{\bm \xi},t_n)$. Fig.\ref{fig} shows the schematic in one-dimensional case.

Since the present scheme is targeting the numerical NS solutions in a resolved dissipative region, thus, the Chapman-Enskog expansion can be employed to approximate the original distribution function. Then, combining with Eq.~\eqref{trans1}, two approximations can be applied to Eq.~\ref{DB2}. First, ${\bar{f}}^+({\bm x}_b,{\bm \xi},t_n)$ is approximated by the first-order Chapman-Enskog expansion of $f$ in the context of the BGK equation, i.e $f\approx g-\tau(g_t+{\bm \xi}\cdot\nabla g)$, and the second approximation is that $ \nabla {\bar{f}}^+({\bm x}_b,{\bm \xi},t_n)\approx \nabla g({\bm x}_b,{\bm \xi},t_n)$, which is consistent with the first-order Chapman-Enskog expansion and includes only the first-order derivatives of the hydrodynamic variables $\rho$, $\bm u$, and $T$. Up to this point, Eq.~\eqref{DB1} can be expressed as
\begin{equation}\label{DB3}
\bar{f}({\bm x}_b,{\bm \xi},t_n+h)={\bar{f}}^+({\bm x}_b,{\bm \xi},t_n)-h{\bm \xi}\cdot\nabla g({\bm x}_b,{\bm \xi},t_n).
\end{equation}
Then based on the compatibility condition and the relation between $f$ and $\bar{f}$, the conservative variables at the cell interface can be obtained,
\begin{equation}\label{DB4}
{\bm W}({\bm x}_b,t_n+h)=\int \bm{\psi} \bar{f}({\bm x}_b,t_n+h)d\Xi,
\end{equation}
from which the equilibrium distribution function $g({\bm x}_b,t_n+h)$ at the cell interface can be obtained. Therefore, based on Eq.~\eqref{trans1} and the obtained equilibrium state, the original distribution function can be extracted from $\bar{f}({\bm x}_b,t_n+h)$,
 \begin{equation}\label{DB5}
{f}({\bm x}_b,t_n+h)=\left(\frac{2\tau}{2\tau+h}-\frac{2\tau-h}{2\tau+h}\tau A-\tau {\bm \xi}\cdot{\bm a}\right)g({\bm x}_b,t_n)+\frac{h}{2\tau+h} g({\bm x}_b,t_n+h),
\end{equation}
where ${\bm a}g=\nabla g,Ag=\partial_t g$. Explicitly, $\bm a$ and $A$ can be expressed as \cite{LC}
\begin{subequations}\label{DB6}
\begin{eqnarray}
{\bm a}=\nabla\ln\rho+\frac{{\nabla {\bm{u}} \cdot ({\bm \xi}  - {\bm{u}})}}{{RT}} + \left( { - \frac{{(D + K)}}{{2}} + \frac{{[{{({\bm \xi}  - {\bm{u}})}^2} + {\bm \eta ^2}]}}{{2RT}}}\right)\nabla \ln T,
\end{eqnarray}
\begin{eqnarray}
A&=&
 -{\bm{\xi}} \cdot \nabla \ln \rho  - \frac{{({\bm{\xi}} - {\bm{u}}) \cdot {\bm{u}} \cdot \nabla {\bm{u}}}}{{RT}} - \frac{{[{{({\bm{\xi}} - {\bm{u}})}^2} + {\bm \eta ^2}]}}{{(D + K)RT}}\nabla \cdot {\bm{u}} \nonumber\\&&- ({\bm{\xi}} - {\bm{u}}) \cdot \nabla \ln T -\left( {\frac{{[{{({\bm{\xi}} - {\bm{u}})}^2} + {\bm \eta ^2}]}}{{RT}} - \frac{{(D + K)}}{2}} \right){\bm{u}} \cdot \nabla \ln T.
\end{eqnarray}
\end{subequations}

Substituting Eq.~\eqref{DB5}  into Eq.~\eqref{fluxh}, the numerical fluxes across the cell interface can be computed, and according to  Eq.~\eqref{update1}, the conservative variables at $t^{n+1}$ can be updated. The above procedure can be repeated in the next time step.


Finally, we come to the Prandtl number fix problem. Due to the fact that all the molecules, regardless of the velocities, have the same particle collision time $\tau$ in the BGK model, the BGK equation gives a fix Prandtl number which is equal to $1$.
In order to fix the Prandtl number to any realistic value, here we use the heat flux modified method proposed in Ref.~\cite{Xu_Prandtl},
\begin{equation}\label{prandtl}
F_{\rho \epsilon}^{new}=F_{\rho \epsilon}+\left(\frac{1}{Pr}-1\right)q,
\end{equation}
where $F_{\rho \epsilon}$ is the energy flux in Eq.~\eqref{flux}, and $q$ is the heat flux,

\begin{equation}\label{heatflux}
q=\int \left(\bm{\xi}-\bm{u}\right)\left(\left(\bm{\xi}-\bm{u}\right)^2+\bm{\eta}^2\right)d\Xi.
\end{equation}

 It is noted that, although sharing a common kinetic origin, there are some distinctive features in the semi-Lagrangian gas-kinetic scheme and the original GKS. First and foremost, different approaches are employed in the construction the flux at the cell interface.  In the original GKS, the flux is evaluated from an integral solution of the BGK equation \cite{Xu_Prandtl}, while in the present scheme the flux evaluation is based on the discrete characteristic solution of the BGK equation that results in a simpler formulation than the original one. Second, unlike the original GKS uses the conservative variables at the previous time step $(t=t_n)$ in the reconstruction of the distribution function, the present scheme uses the ``instaneous"  half-time variables $(t=t_n+h)$.

\section{Numerical Validations}
\label{numerical}
In this section, the proposed scheme is validated by simulating several test problems, including the thermal Couette flow, the thermal Poiseuille flow, the shock structure problem, and the laminar flow past a flat plate. In all of the numerical simulations, the collision time $\tau$ is determined by $\tau=\mu/p$, where $\mu$ is the dynamical viscosity and $p=\rho RT$ is the pressure; The time step is determined by the CFL condition, i.e., $\Delta t=\eta{\Delta x}_{min}/\left(C_{max}+c_s\right)$, where $\eta$ is the CFL number and is set to be $0.5$ unless otherwise stated, ${\Delta x}_{min}$ is the minimum mesh spacing, $C_{max}$ is the maximum velocity of flow, and $c_s=\sqrt{\gamma RT}$ is the speed of sound, $\gamma$ is the special heat ratio; The gas is assumed to be monotonic such that $\gamma=5/3$ unless otherwise stated; The central difference is applied to approximate the gradients of the conservative variables in spatial space, which in theory yields present scheme second-accurate in space.

\subsection{Thermal Couette flow }
\begin{figure}
\centering
\includegraphics[width=0.6\textwidth]{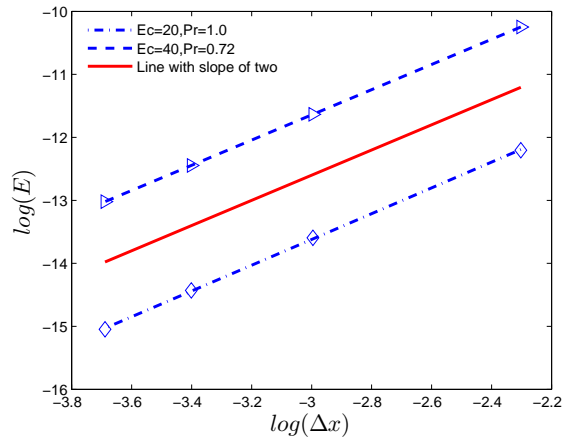}
\caption{Numerical errors versus grid size with different $\text{Ec}$ and $\text{Pr}$. }\label{error}
\end{figure}

\begin{figure}
\centering
\includegraphics[width=0.6\textwidth]{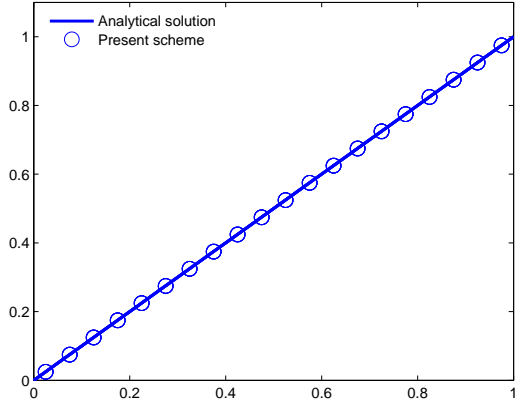}
\caption{The horizontal velocity distribution along the vertical direction of the channel with grid spacing $\Delta x=0.05$.}\label{velocityco}
\end{figure}

 The thermal Couette flow is a standard test case, which has an analytical solution\cite{Xu_Prandtl,HE,SHY}. The problem considered is  an incompressible and viscous fluid between two infinite parallel flat plates with a distance $H$, the upper plate fixed at temperature $T_1$ moves at a speed $U$ in the horizontal direction, and the lower plate fixed at temperature $T_0\left(T_1>T0\right)$ is stationary. In the simulations, we set $H=1$, $U=1$, $T_1=1$, $T_0=0$, and the Mach number $\text{Ma}=U/c_s=0.1$ for the near incompressible limit;  No-slip boundary conditions are applied to both the upper and lower plates \cite{Xu_Prandtl}, and periodic boundary conditions to the inlet and outlet of the channel. Under the assumption of constant viscosity and heat conduction coefficients, the temperature and velocity can be obtained analytically as \cite{Xu_Prandtl},
\begin{equation}\label{analysis}
\frac{T-T_0}{T_1-T_0}=\frac{y}{H}+\frac{\text{Pr}\text{Ec}}{2}\frac{y}{H}\left(1-\frac{y}{H}\right),
\end{equation}
\begin{equation}\label{analysisv}
 u=U\frac{y}{H},
\end{equation}
where $y$ is the height relative to the lower plate, $\text{Pr}$ is the Prandtl number, $\text{Ec}$ is the Eckert number defined as $\text{Ec}={U}^2/C_p(T_1-T_0)$, and $C_p$ is the special heat ratio at constant pressure.

In order to evaluate the accuracy of the present scheme, the relative global errors of temperature  with various mesh resolutions are measured, where the relative global error is defined as
\begin{equation}\label{err}
E(T)=\Sigma \frac{\parallel T-T_e\parallel_1}{\parallel T_e\parallel_1},
\end{equation}
in which $T$ is the temperature obtained by the present scheme, $T_e$ is the analytical solution. In the simulation, we set $\tau=\Delta t$ and the grid spacing varies from $1/10$ to $1/40$ in the vertical direction.
 As showed in Fig.~\ref{error}, the slopes of the fitting lines of relative errors for the temperature field with $\text{Ec}=20$, $\text{Pr}=1.0$ and  $\text{Pr}=0.72$, $\text{Ec}=40$ are equal to $2.04$ and $1.99$, respectively, confirming that the new scheme is of second-order accuracy in space. Under the same initial condition and grid spacing of $1/20$ in the vertical direction, the horizontal velocity $u$ distribution along the vertical direction of the channel, as well as the analytical solution are shown in Fig.~\ref{velocityco}. It is observed that the results agree well with the analytical solution.
\begin{figure}
\centering
\includegraphics[width=0.6\textwidth]{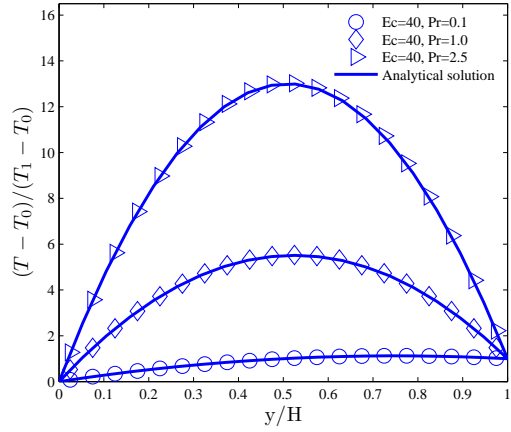}
\caption{Non-dimensional temperature distributions with different Prandtl number at $\text{Ec}=40$. }\label{Ec=40}
\end{figure}
\begin{figure}
\centering
\includegraphics[width=0.6\textwidth]{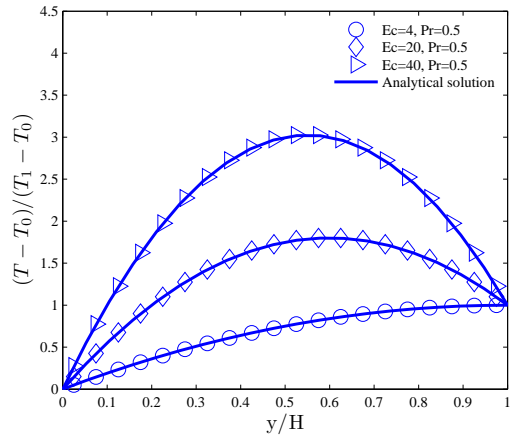}
\caption{ Non-dimensional temperature distributions with different Eckert number at $\text{Pr}=0.5$.}\label{Pr=0.5}
\end{figure}

In addition, we also test the current model at a wide range of the $\text{PrEc}$ which represents the ratio between the viscous dissipation and the heat conduction. For small values of $\text{PrEc}$, temperature varies almost linearly along the direction perpendicular to the plates, which means that the effect of the viscous heat dissipation is rather weak compared with heat conduction. As $\text{PrEc}$ increases, the viscous heat dissipation turns to be dominant, and the temperature profile deviates from the linear distribution.
Fig.~\ref{Ec=40} shows the non-dimensional temperature distributions at different Prandtl numbers with $\text{Ec}=40$, and Fig.~\ref{Pr=0.5} depicts the non-dimensional temperature distributions at different Eckert numbers with $\text{Pr}=0.5$, from which it can be seen that the numerical results of the present model are in an excellent agreement with the exact solutions. As is shown , the present scheme can accurately describe the viscous and heat conducting flow with a wide range of  $\text{PrEc}$ from $2.0$ to $100$.

\subsection{Thermal Poiseuille flow}
The thermal Poiseuille flow is another crucial test problem \cite{SHY,DF}. Unlike the thermal Couette flow, in this case, the parallel plates are all stationary and the flow between the plates is driven by a constant force $G$.  The temperatures of the top and bottom plates are kept at $T_1$ and $T_0$.  Under the incompressible condition, the analytical solutions are given by \cite{DF}

\begin{equation}\label{analysis_pov}
u=\frac{Gy}{2\mu H}(1-\frac{y}{H})
\end{equation}
\begin{equation}\label{analysispo}
\frac{T-T_0}{T_1-T_0}=\frac{y}{H}+\frac{\text{PrEc}}{3}\left[1-\left(1-\frac{2y}{H}\right)^4\right],
\end{equation}
where $\mu=\sqrt{H^3G/\text{Re}}$ is the shear viscosity, $y$ is the height relative to the lower plate, $H$ is the height of the channel. In our simulations, $100$ grids are employed along the vertical direction;  No-slip boundary conditions are applied to the two plates \cite{Xu_Prandtl}, and periodic boundary conditions to the inlet and outlet of the channel. The external force is realized by the operator splitting method \cite{xu1999}.

\begin{figure}
\centering
\includegraphics[width=0.6\textwidth]{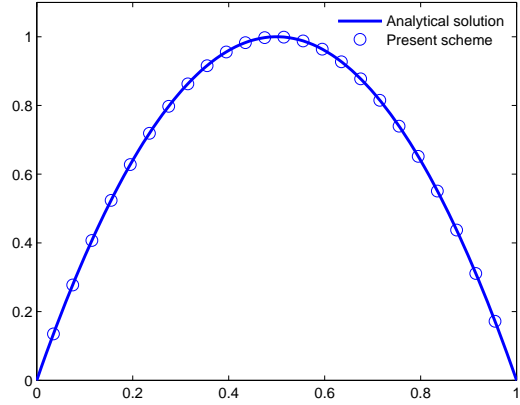}
\caption{The horizontal velocity distribution along the vertical direction of the channel in the Poiseuille flow at $Re=100$ and the grid spacing $\Delta x=0.01$ .}\label{velocitypo}
\end{figure}

\begin{figure}
\centering
\includegraphics[width=0.6\textwidth]{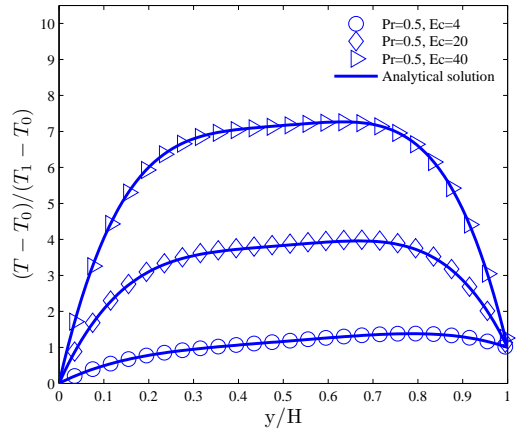}
\caption{ Non-dimensional temperature distributions with different Eckert numbers at $\text{Pr}=0.5$.
}\label{tempr}
\end{figure}

\begin{figure}
\centering
\includegraphics[width=0.5\textwidth]{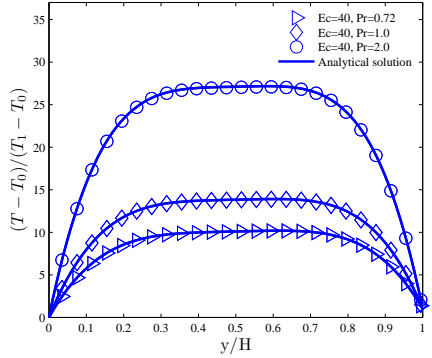}
\caption{ Non-dimensional temperature distributions with different Prandtl numbers at $\text{Ec}=40$.}\label{temec}
\end{figure}

 The Poiseuille flow is characterized by the Reynolds number $\text{Re}=\rho H U_0/\mu$, where $U_0=\rho G H^2/8\mu$. In our simulations, the Reynolds number $\text{Re}$ is set to be $100$;  The Mach number is set to be $0.1$ so that the flow is near incompressible. We conduct a set of simulations at different values of $\text{Pr}$ and $\text{Ec}$. Fig.~\ref{velocitypo} shows the distribution of the horizontal velocity along the vertical direction of the channel, also included is the analytical solution.  The non-dimensional temperature profiles for different Eckert numbers at $\text{Pr}=0.5$ are presented in Fig.\ref{tempr}, and the non-dimensional temperature profiles for different Prandtl numbers at $\text{Ec}=40$ are shown in Fig.\ref{temec}. It is clearly seen from these figures that the numerical results are in excellent agreement with the analytical solutions in both flow and temperature fields.

\subsection{Navier-Stokers Shock structure}
We now test the proposed scheme by calculating the Navier-Stokes shock structure. Although the Navier-Stokes solutions do not give the
physically realistic shock wave profile in high Mach number case, it is still a useful problem in establishing and testing a valid solver
 for the Navier-Stokes equations. Even though the shock structure is well resolved in this case, due to the highly non-equilibrium inside
 the shock layer, its accurate calculation bears large requirement on the accuracy and robustness of the numerical method. The profile of a
normal shock structure represents a good test for viscous flow solvers.


\begin{figure}
\centering
\subfigure{\includegraphics[width=0.45\textwidth]{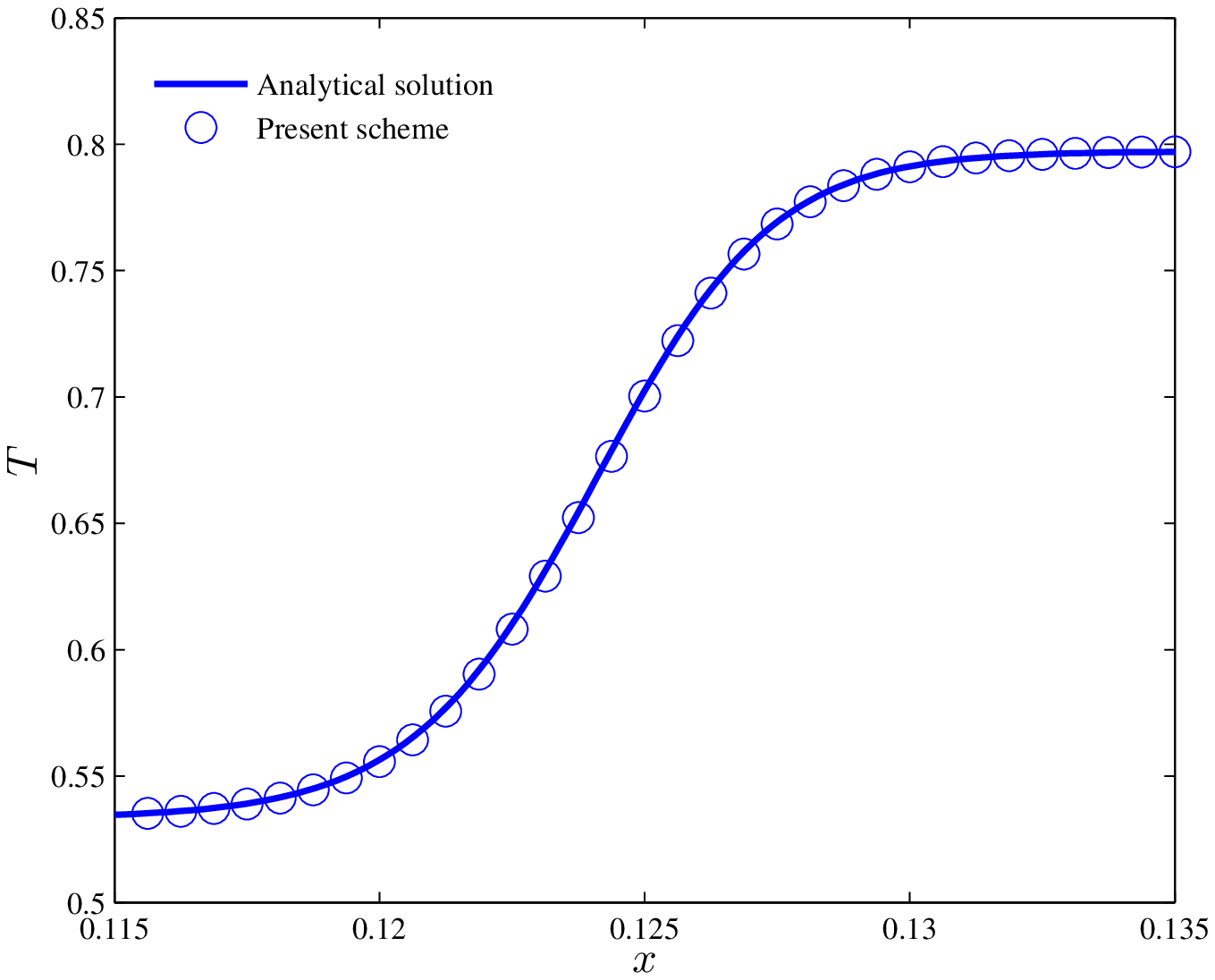}}\hfill
\subfigure{\includegraphics[width=0.45\textwidth]{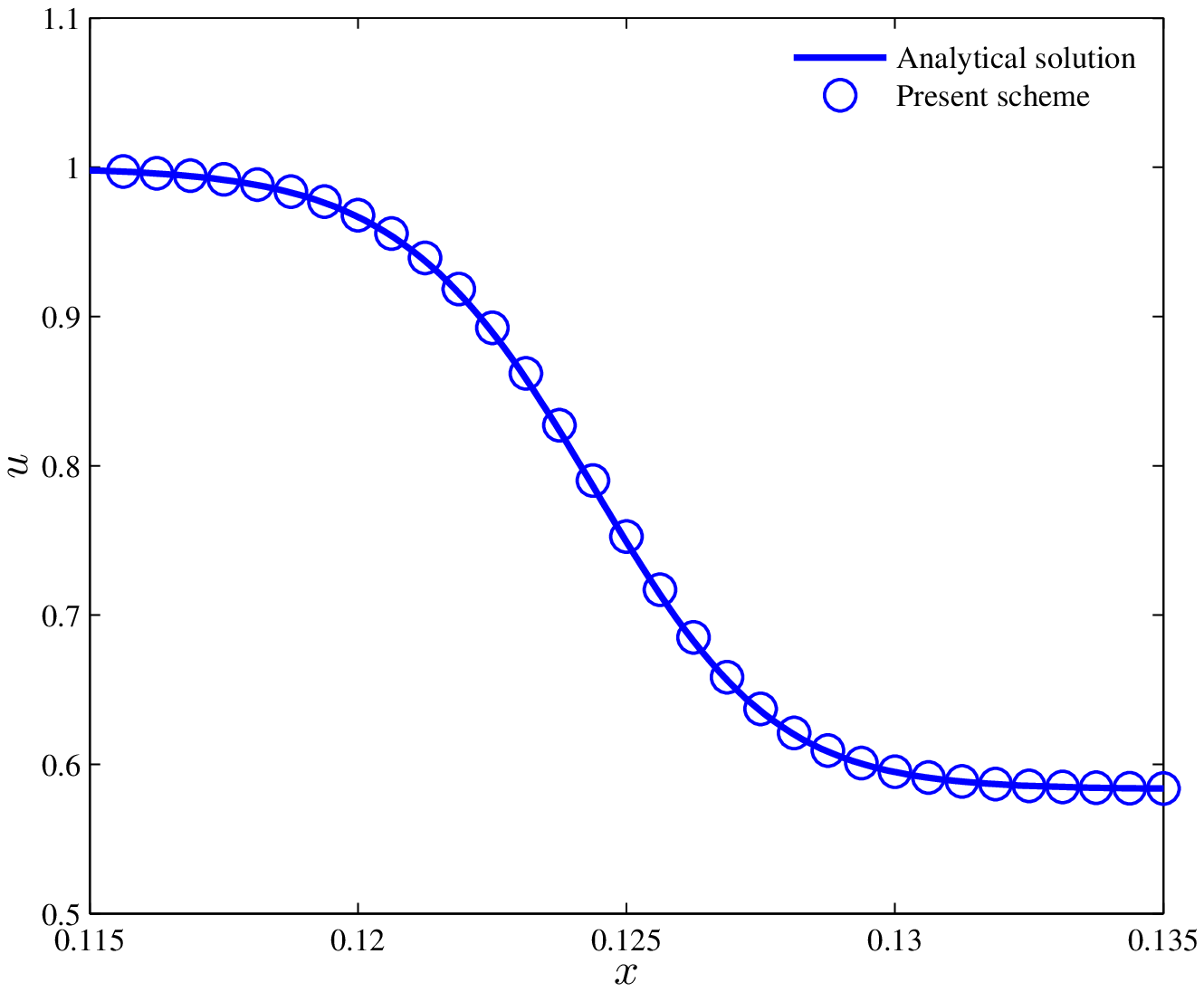}}
\caption{Temperature and velocity distributions inside the shock structure for a monatomic gas with Ma=1.5, Pr=1.0.}\label{tem1}
\end{figure}
In our simulation, the shock structure for a monotonic gas with $\gamma =5/3$ and a viscosity coefficient $\mu \sim T^{0.8}$ is considered.
Two Prandtl numbers, $\text{Pr}=1.0$ and $\text{Pr}=2/3$, are considered with a fixed Mach number $\text{Ma}=1.5$. The dynamic viscosity coefficient at the upstream keeps a constant $\mu_{-\infty}=0.0005$. The referenced ``exact" solution is obtained by directly integrating the steady state  Navier-Stokes equation \cite{Xu_Prandtl}. The numerical solution depends closely on the scales of the mesh resolution and physical flow structure, and it requires a fine mesh resolution to resolve the wave structure in this case. Hence, in our calculations, the mesh size is chosen to be $\triangle x=1/1600$.

\begin{figure}
\centering
\subfigure{\includegraphics[width=0.45\textwidth]{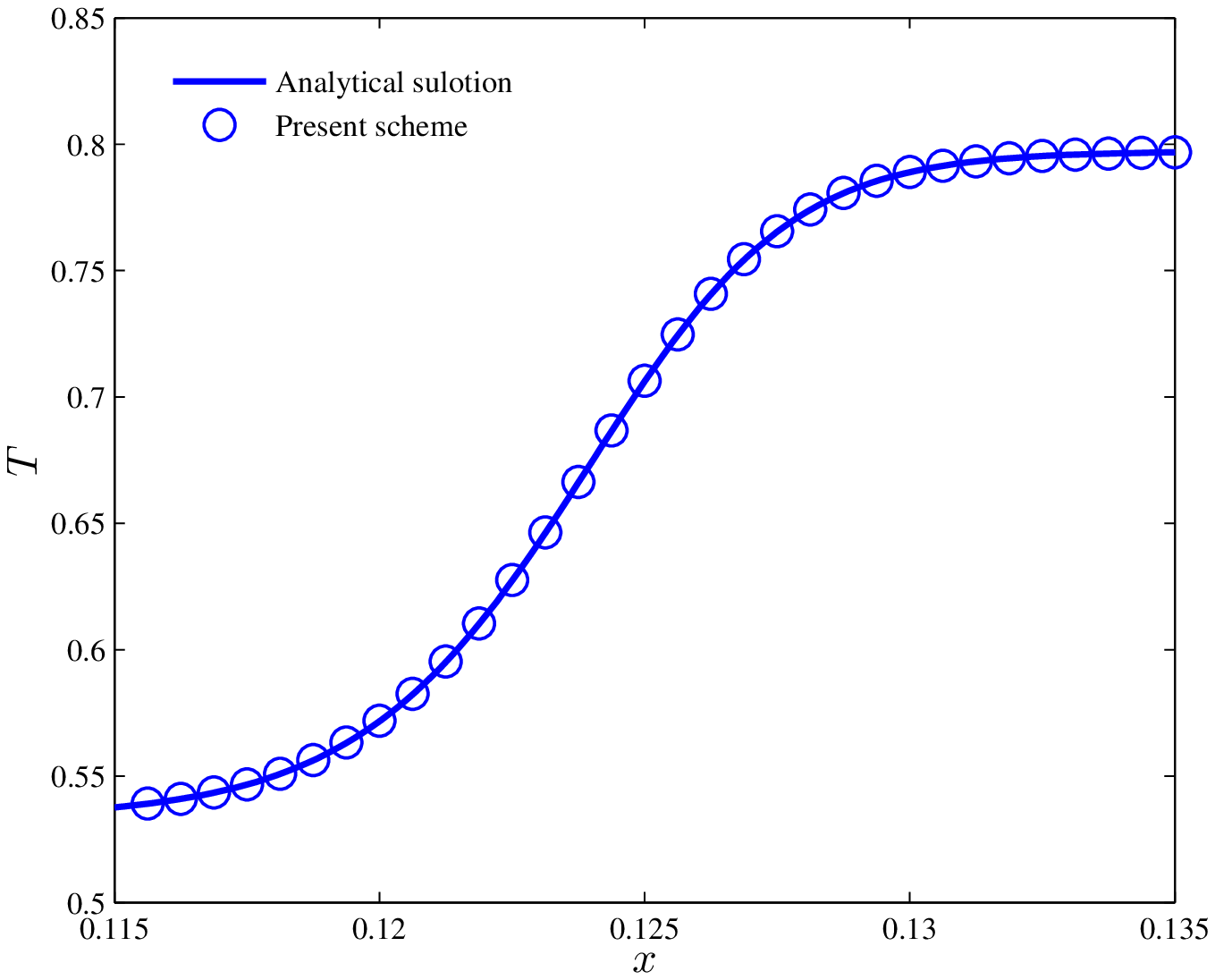}}\hfill
\subfigure{\includegraphics[width=0.45\textwidth]{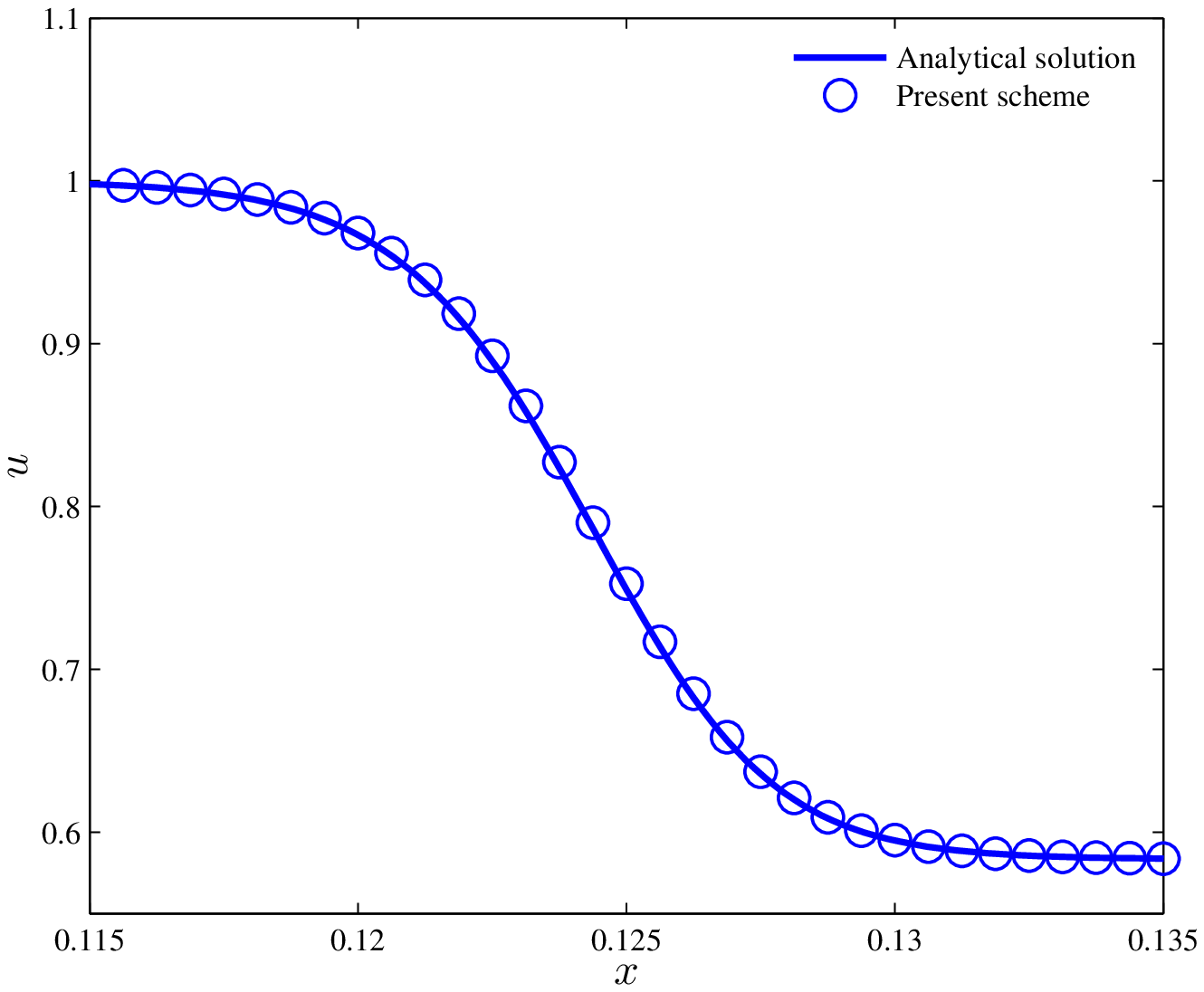}}
\caption{Temperature and velocity distributions inside the shock structure for a monatomic gas with Ma=1.5, Pr=2/3.}\label{tem23}
\end{figure}

The profiles of temperature $T$ and  velocity $u$ calculated by the SLGKS across the shock layer  for $\text{Pr}=1.0$ and $\text{Pr}=2/3$ are shown in Fig.~\ref{tem1} and Fig.~\ref{tem23}, respectively. It is clearly observed that the shock structure obtained from the SLGKS matches with the ``exact" solutions excellently.

\subsection{Laminar flow past a flat plate}
\begin{figure}
\centering
\includegraphics[width=0.8\textwidth]{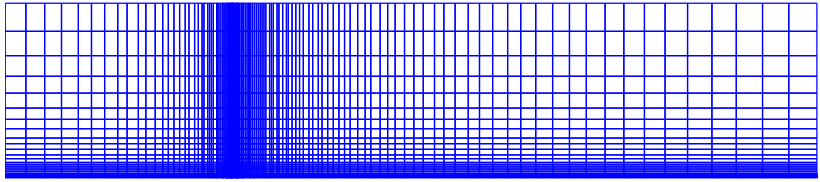}
\caption{ Non-uniform mesh  used for the laminar flow past a flat plate.}\label{mesh}
\end{figure}

\begin{figure}
\centering
\includegraphics[width=0.8\textwidth]{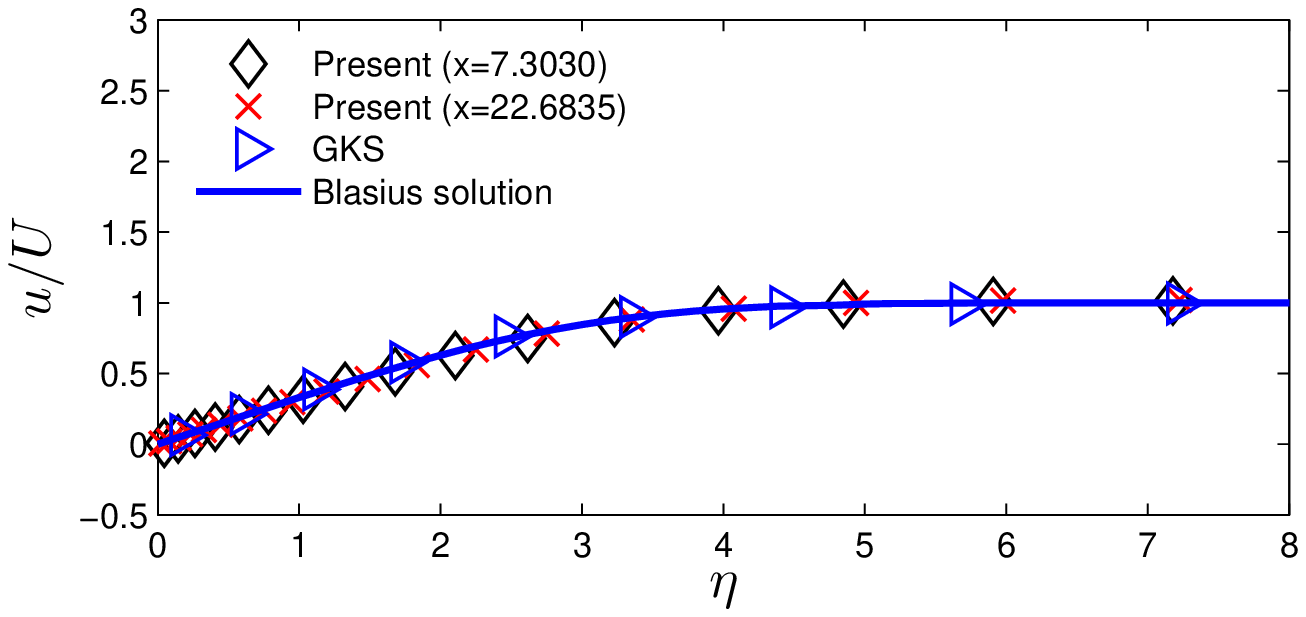}
\caption{Comparison of the U-velocity profiles in the $x-$direction  at different locations obtained using SLGKS at $\text{Re}=10^4$ with GKS and Blasius solution.}\label{Ree4}
\end{figure}

\begin{figure}
\centering
\includegraphics[width=0.8\textwidth]{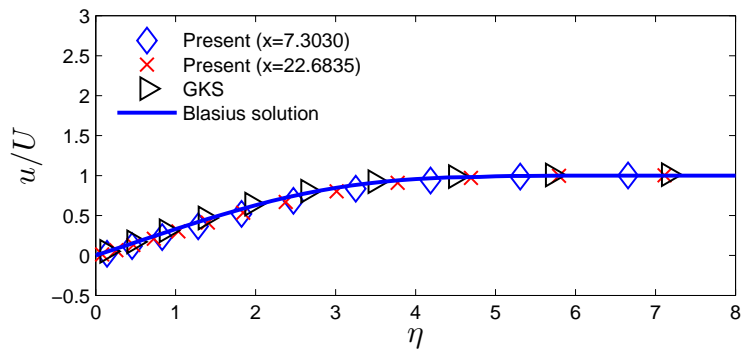}
\caption{Comparison of the U-velocity profiles in the $x-$direction  at different locations obtained using SLGKS at $\text{Re}=10^5$ with GKS and Blasius solution.}\label{Ree5}
\end{figure}

The laminar boundary layer over a flat plate is a typical problem to test the numerical dissipation of a new scheme. The problem considered is the laminar flow with a constant horizontal velocity $U$ over a semi-infinite flat plate , which has an exact self-similar Blasius solution. In this test case, the Reynolds number is defined as
\begin{equation}\label{Reynold}
\text{Re}=\frac{UL}{\nu},
\end{equation}
where $L$ is the length of the plate, $\nu$ is the kinematic viscosity. In order to capture the boundary layer accurately, a non-uniform mesh with $120\times30$ grid points is adopted in our simulation, as sketched in Fig.~\ref{mesh}, where the mesh resolution is varying according to the local accuracy requirement. In the simulations, the flat plate is placed from  $x=0$ to the right side of the mesh; The initial inflow boundary condition at the left boundary is \cite{Xu_Prandtl}
\begin{equation}\label{initial}
\left(\rho, U, V, p\right)=(1, 3, 0, 9/\gamma \text{Ma}^2),
\end{equation}
where the Mach number $\text{Ma}$ is set to be $0.15$ in the simulation; No-slip boundary condition is imposed on the flat plate, the non-reflecting boundary condition, based on the one dimensional Riemann invariants normal to the grid, is used at the upper boundary, and the simple extrapolation of the conservation variables are used on the right boundary.
Fig.~\ref{Ree4} and Fig.~\ref{Ree5} show the $x-$ velocity profiles at different locations along the vertical direction insider the boundary layer at $\text{Re}=10^4$ and $\text{Re}=10^5$, respectively. For comparison, the GKS and exact Blasius solutions are also included. It can be found that the results given by the new scheme agree well with the exact Blasius solution and results of GKS, which indicate that the new scheme exhibits a low numerical dissipation. From these test cases, we clearly observe that the semi-Lagrangian gas kinetic scheme solves the NS equations accurately.



\section{Conclusions}
\label{con}
In this paper, a semi-Lagrangian kinetic approach for smooth flows based on the BGK equation is derived. With the use of the Lagrangian perspective, a simple formulation of the distribution function at the cell interface is derived and the particle transport and collision are coupled together in the evaluation of the flux at the cell interface. As a result, the scheme exhibits low numerical dissipation and the time step is solely determined by the CFL condition. The numerical simulations show that the scheme exhibits second-order spatial accuracy, and can approximate the Navier-Stokes solutions accurately. In conclusion, the SLGKS in a simple formulation is an alternative feasible solver for Navier-Stokers equations in smooth region.

We would like to emphasize that the present scheme mainly focuses on the smooth flows, further development for flows with discontinuities will be presented in subsequent papers.
\section*{Acknowledgements}
\label{ack}
This study is financially supported by the National Natural Science Foundation of China
(Grant No. 51125024).

\end{document}